\documentclass[twocolumn,aip,apl,reprint,showpacs,preprintnumbers,amsmath,amssymb,floatfix,groupedaddress]{revtex4-1}
\usepackage{graphicx,CJK,times}
\usepackage{amsmath}
\usepackage{amssymb}
\usepackage{color}

\begin{document}

\title{Electrical conductivity of multi-walled carbon nanotubes-SU8 epoxy composites} 

\author{Claudio Grimaldi} 
\author{Marijana Mioni\'c}\altaffiliation[Current address: ]{Powder Technology Laboratory,
Ecole Polytechnique F\'ed\'erale de Lausanne, CH-1015 Lausanne, Switzerland}
\author{Richard Gaal}
\author{L\'aszl\'o Forr\'o} 
\author{Arnaud Magrez}
\affiliation{Laboratory of Physics of Complex Matter, Ecole Polytechnique F\'ed\'erale de Lausanne, CH-1015 Lausanne, Switzerland}  

\begin{abstract}
We have characterized the electrical conductivity of the composite which consists of multi-walled carbon nanotubes 
dispersed in SU8 epoxy resin. Depending on the processing conditions of the epoxy (ranging from non-polymerized  
to cross-linked) we obtained tunneling and percolating-like regimes of the electrical conductivity of the composites.
We interpret the observed qualitative change of the conductivity behavior in terms of reduced separation between
the nanotubes induced by polymerization of the epoxy matrix. 
\end{abstract}
%\pacs{64.60.ah, 82.70.Dd, 82.70.Gg, 73.40.Gk}
\maketitle

%\section{introduction}
%\label{intro}

Carbon nanotubes, 20 years after their discovery are in the stage where the focus is more and more on their 
applications. The physical properties which are beneficial for applications are good electrical conductivity ($\sigma$), 
exceptionally high thermal conductivity, and their high mechanical strength. These properties are preferentially used 
in composites, where carbon nanotube fillers can provide the missing property of the matrix, for example the electrical 
conductivity. The metallic carbon nanotubes (CNTs) added in concentrations beyond the percolation threshold to the insulating 
matrix can give an antistatic composite,\cite{Byrne2010} transparent conducting electrode,\cite{Brennan2011} 
or other large area conducting materials.\cite{Lota2011}

\begin{figure}[b]
\begin{center}
\includegraphics[scale=1.0,clip=true]{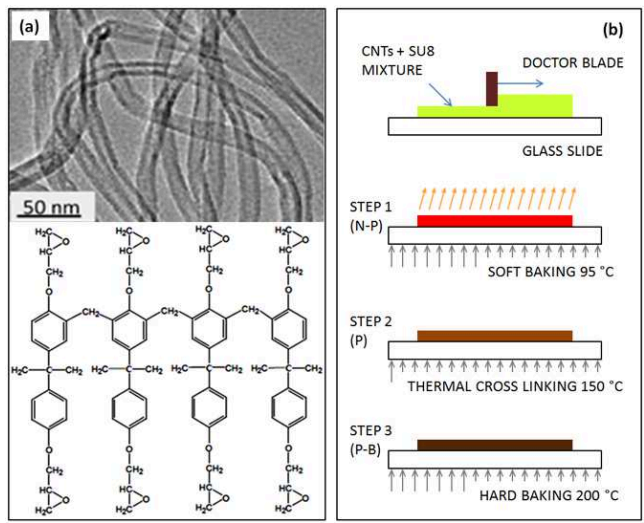}
\caption{(a) Upper panel: Transmission Electron Microscopy image of an assembly of CNTs prepared by a CVD method. 
The average diameter was $13.3$ nm, while the length was approximately $10$ $\mu$m; Lower panel: schematic representation 
of the SU8 epoxy oligomer. 
(b) The preparation steps of the CNTs-SU8 composites for electrical conductivity measurements. }
\label{fig1}
\end{center}
\end{figure}

In composite materials, one main issue is the possibility of combining the properties of the filler with those of the 
matrix. In this respect, SU8 matrix, which is a well-established engineering material for micro- and nano-fabrication, offers
wide range opportunities. SU8 is an epoxy-based, negative-tone, UV-sensitive photoresist which
besides SU8 resin, contains an organic solvent and a photoinitiator (PI) to provide cross-linking of the oligomers
(see a schematic representation of the SU8 epoxy in Fig.~\ref{fig1}).\cite{USPatents} 

One of the great advantages of SU8 is that it can allows fabrication of high aspect ratio three-dimensional structures 
in a broad range of thicknesses and hence it is often used in nano- and macrometer sized devices.\cite{Teh2005} 
The disadvantages of SU8 are that it is an electrical insulator, it is brittle and it has a low thermal
conductivity. It would be desirable not to have all these drawbacks, so numerous fillers materials were used to prepare
SU8-based composites.\cite{Jiquet,Xu2002,Zhang2003} 
 
Here we report the electrical conduction of multi-walled carbon nanotubes-SU8 composite 
materials (hereafter CNTs-SU8) for a broad concentration range of well-dispersed CNTs. 
The CNTs-SU8 inks were prepared with and without PI, giving composites with and without cross-linked matrix, respectively.
From the theoretical point of view, these two cases are equivalent to 
study the conduction mechanism of an assembly of CNTs in a solid and in a ``liquid'' matrix. Interestingly, beside
the effects on the conductivity level, the functional dependence of $\sigma$ on the CNTs' content noticeably differs 
for these two cases.

\begin{figure}[t]
\begin{center}
\includegraphics[scale=1.0,clip=true]{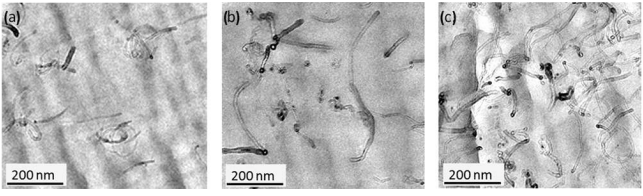}
\caption{Transmission Electron Microscopy images of microtome slices of CNT-SU8 composites for different 
volume fraction $\phi$ of CNTs: a) $\phi=0.0023$,  
b) $\phi=0.0046$, and c) $\phi=0.008$. The images illustrate the good dispersion of the CNTs in the SU8 matrix.}
\label{fig2}
\end{center}
\end{figure}

%\section{experimental}
%\label{experimental}

%%%%%%%%%%%%%%%%%%%%%%%%%%%%%%%%%%%%%%%%%%%%%%%%%%%%%%%%%%%%%%
\begin{figure*}[t]
\begin{center}
\includegraphics[scale=1.0,clip=true]{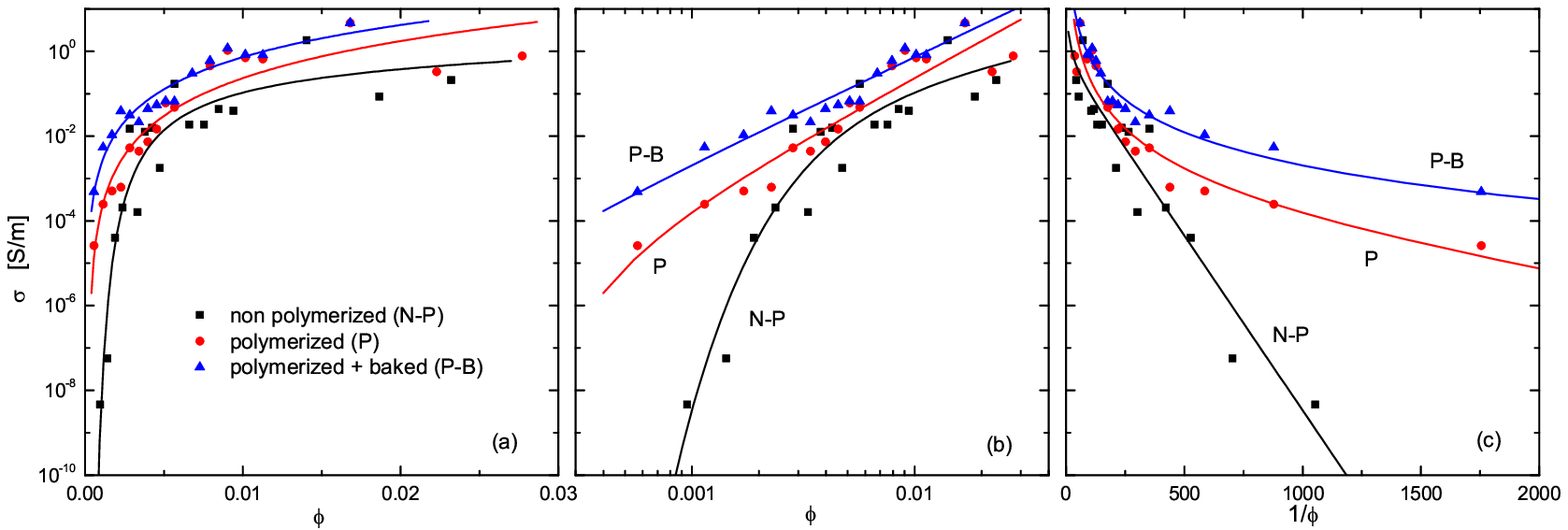}
\caption{(a) Results of conductivity measurements (symbols) for CNT-SU8 composites with non-polymerized epoxy (N-P), 
with polymerized (P), and with polymerized hard-baked sample (P-B). The solid lines are least square fits of the conductivity
data to the theory. 
(b) Conductivity plotted on a log-log scale: the P and P-B composites display a power-law behavior as $\phi\rightarrow 0$.
(c) Conductivity on a log scale plotted as a function of $1/\phi$: the N-P composites follows the tunneling behavior 
with $D^2/\xi L=0.019$.}
\label{fig3}
\end{center}
\end{figure*}
%%%%%%%%%%%%%%%%%%%%%%%%%%%%%%%%%%%%%%%%%%%%%%%%%%%%%%%%%%%%%%

Multiwalled CNTs were synthesized by the chemical vapor decomposition method (CVD) at $640$ $^\circ$C using 
Fe-Co catalytic particles supported by CaCO$_3$. Afterwards, CNTs were purified in order to remove the catalytic particles 
and the supporting material. A typical TEM image of CNTs is shown in the upper panel of Fig.~\ref{fig1}(a). 
The average CNTs' diameter distribution was $13.3$ nm, while the typical length was approximately $10$ $\mu$m.\cite{Mionicthesis}
CNTs were dispersed in the SU8 by sonication in the presence of surfactant.\cite{Mionic}  
In the last step PI was added for the sample foreseen for polymerization.\cite{PatentLPCM} 
Afterwards, the ink was spread by doctor blading on a clean glass slide. 
In this study, we had three sets of composite samples. 
By soft baking of inks on glass slide at $95$ $^\circ$C, the solvent was evaporated (step 1). At this point we obtained samples 
(the first set) of composites with non-polymerized matrix. The matrix of composites was thermally cross-linked by baking at 
$150$ $^\circ$C,\cite{Ong2006} what gave composite samples (the second set) with polymerized SU8. 
Subsequent hard bake (an optional step in the standard SU8 processing) at $200$ $^\circ$C provides the third set of the samples 
studied in this work.
The schematic representation of the sample preparation is shown in Fig.~\ref{fig1}(b). Figure \ref{fig2} shows
TEM images of microtome slices of CNT-SU8 composites.
From each step a sample of characteristic planar size of $2\times 1$ cm$^2$ was tailored and prepared for 4-contact 
resistivity measurements. 
The composites contained from $x = 0.1\%$ to $x = 5\%$ of CNTs in weight with respect to the weight of SU8.  
Up to $19$ different CNTs concentrations were prepared and measured.

%\section{results and discussion}
%\label{results}

In Fig.~\ref{fig3} the results of 4-point electrical conductivity ($\sigma$) measurements are shown for the CNTs-SU8 composites as a 
function of the volume fraction of CNTs for three different stages of CNTs-SU8 composites with respect to the cross-linking 
of the polymer matrix: non-polymerized (denoted hereafter as N-P), polymerized (denoted hereafter as P), and polymerized and 
baked (denoted hereafter as P-B). On the sketch of Fig.~\ref{fig1}(b) these stages correspond to the processing steps 1, 2 and 3. 
In order to facilitate comparison with theory, in Fig.~\ref{fig3} we report $\sigma$ as a function of the CNT volume fraction $\phi$, 
defined as $\phi=\rho_\textrm{SU8}x/(\rho_\textrm{CNT}+\rho_\textrm{SU8}x)$, where $x$ is the weight ratio of CNTs and SU8,  
$\rho_\textrm{SU8}=0.998$ g/cm$^3$ and $1.218$ g/cm$^3$ are the mass densities of non-polymerized and polymerized SU8,
respectively,\cite{Feng2002} and $\rho_\textrm{CNT}=2.1$ g/cm$^3$ is the mass density of CNT.

Two main features are distinguishable from Fig.~\ref{fig3}. 
First, the conductivity values at low $\phi$ for composites P and P-B are larger than those for the composite N-P at similar volume 
fractions. Second, as it can be inferred from Figs.~\ref{fig3}(b) and (c), the behavior of $\sigma$ 
as $\phi$ goes to zero for the N-P is qualitatively different from that displayed by composites P and P-B. In particular, the 
conductivity data for the two latter composites follow approximately a straight line when plotted in a double-logarithmic scale,
as shown in Fig.~\ref{fig3}(b).
This trend is consistent with a power-law behavior of the form $\sigma\sim (\phi-\phi_c)^t$, with a critical concentration $\phi_c$ close 
to zero, which suggests a percolation-like mechanism of transport.\cite{Stauffer1992} On the contrary, as shown in Fig.~\ref{fig3}(b),
the conductivity prior to polymerization (N-P samples) drops faster than a power-law as $\phi\rightarrow 0$.
It turns out that in the case of N-P composites the measured conductivity data follow approximately a straight 
line when plotted as $\ln(\sigma)$ versus $1/\phi$, Fig~\ref{fig3}(c), which is consistent with a tunneling mechanism of 
transport between conducting CNTs dispersed homogeneously in the matrix.\cite{Ambrosetti2010} 

The conductivity data shown in Fig.~\ref{fig3} suggest that polymerization-caused change in the composite structure 
is responsible for the change in the conductivity behavior observed for the composites with polymerized and non-polymerized matrix. 
In particular, this structural effect appears to favor the inter-tube electrical connectedness and to simultaneously 
turn the functional dependence of $\sigma$ from the tunneling-type in N-P composites to the percolation-like in P
and P-B composites.

A possible structural mechanism capable of explaining this behavior could be identified in the reduced CNT separation
induced by the polymerization of the SU8 matrix. Indeed, while the $\ln(\sigma)\propto -1/\phi$ behavior is expected for
homogenous dispersions of nanotubes,\cite{Ambrosetti2010} a percolation-like conductivity arises when current flows through 
clusters of conducting objects in close contact. 

To show in more details how reduced inter-tube separations could account for the tunneling-to-percolation crossover
shown in Fig.~\ref{fig3}, we model the CNT-SU8 system as given by isotropically oriented tubes of length $L$ and diameter 
$D\ll L$ dispersed in an insulating continuous matrix.
The inter-tube electrical connectedness is established by tunneling processes with conductance $g(r_{ij})=g_0\exp[-2(r_{ij}-D)/\xi]$,
where $r_{ij}$ is the distance between the centerlines of two tubes $i$ and $j$, $g_0$ is the contact conductance, and $\xi$ is the tunneling decay length.
The overall conductivity of the system can be estimated by using the effective medium approximation (EMA) which
replaces each tunneling conductance $g(r_{ij})$ by an effective conductance independent of $r_{ij}$. 
By requiring equivalence between the original tunneling network and the effective one, the resulting EMA conductance
$g^*$ in units of $g_0$ is obtained by the solution of the following equation:\cite{Grimaldi2011}  
\begin{equation}
\label{EMA1}
\frac{2L}{D^2}\phi\int_D^\infty \! dr\,\frac{g_2(r)}{g^*\exp [2(r-D)/\xi ]+1}=1,
\end{equation}
where the radial distribution function $g_2(r)$ is proportional to the probability density of finding a tube at distance 
$r$ from a given tube.\cite{note1} 

Since in Eq.~\eqref{EMA1} the structural characteristics of the CNT dispersions are entirely contained in $g_2(r)$, a 
simple way to simulate the effects of reduced mean tube separations is to use a radial 
distribution function with enhanced population of tubes at close distances. If we neglect details and retain only the gross 
features, a simple model is thus $g_2 (r)=a\delta (r-D)+1$ for $r\ge D$, where $\delta (r-D)$ 
is a Dirac-delta function centered at $D$ and $a$ is a phenomenological parameter measuring the ``stickiness" 
between the nanotubes.\cite{Nigro2012}  
When this form for $g_2(r)$ is inserted in Eq.~\eqref{EMA1} we find:
\begin{equation}
\label{EMA2}
\frac{\chi\phi}{1+g^{*1/t}} +\frac{L\xi}{D^2}\phi\ln\left(\frac{1+g^*}{g^*}\right)=1,
\end{equation}
where $\chi=2La/D^2$. To better describe the percolation behavior, in the denominator of the first term of 
Eq.~\eqref{EMA2} we have replaced $1+g^*$ with $1+g^{*1/t}$, where $t$ is the percolation transport exponent.  
In this way, when the first term in the left hand-side of Eq.~\eqref{EMA2} dominates over the second one, the 
solution reduces to: $g^*=(\phi-1/\chi)^t$, i.e., the conductivity (which is proportional to $g^*$) follows a 
percolation-like behavior with critical concentration $1/\chi$ and transport exponent $t$. 
Conversely, when $\chi = 0$, or when $g^*\ll 1$, the second term of Eq.~\eqref{EMA2} becomes dominant and the transport becomes 
of tunneling type:
\begin{equation}
\label{EMA3}
g^*\approx\exp\left(-\frac{D^2}{\xi L\phi}\right).
\end{equation}
The above expression can be derived also by the method of the critical path approximation, according to which
the conductivity is proportional to $\exp(-2\delta_c/\xi)$,\cite{Ambegaokar1971} where $\delta_c=D^2/2L\phi$ is the 
critical connectedness distance for homogeneous dispersions of slender tubes.\cite{Otten2009,Chatterjee2010}
Percolation-like and tunneling-like regimes can thus be both obtained from Eq.~\eqref{EMA2} as limiting behaviors
governed by the mean inter-particle separation.

In Fig.~\ref{fig3} we show by solid lines the EMA conductivity $\sigma=\sigma_0 g^*$, where $g^*$ is the solution of 
Eq.~\eqref{EMA2} with the parameters $\chi$, $t$ and $\sigma_0$ obtained from the least square fits to the experimental data.
The fitting parameter $\sigma_0$ accounts for the contribution of the contact conductance ($g_0$) to the conductivity.
For the case N-P, we have set $\chi = 0$ (i.e., ``stickiness" parameter $a$ is zero) 
and $D^2/\xi L=0.019$, which reproduces the tunneling behavior of Eq.~\eqref{EMA3}. By using $D\approx 13.3$ nm and 
$L\approx 10$ $\mu$m, we find thus $\xi\approx 0.93$ nm, which is a quite reasonable values for the tunneling decay length. 
For the two polymerized cases (P and P-B), the best fits are obtained for $\chi = 4062$ and $t = 2.86$ for the P 
composites and $\chi = 22500$ and $t = 2.51$ for the hard-backed system P-B, with negligible dependence on $D^2/\xi L$ in the 
explored concentration range. 
Considering that $L/D\approx 750$, the ``stickiness" parameter thus increases from $a/D\approx 2.7$ to 
$a/D\approx 15$ for the P and P-B composites respectively, which indicates an increase in the population of nanotube 
``at contact'' upon hard-backing.  
We note, furthermore, that the values of the transport exponent $t$ are comprised between $t\simeq 2$ and $t=3$, 
which represent percolation in three dimensions and in mean-field regime, respectively.\cite{Stauffer1992,Straley1977} 
Percolation of slender nanotubes is considered to belong to the mean-field universality-class.

It is worth pointing out that the theoretical interpretation of the conductivity data of Fig.~\ref{fig3}
is based on a model of straight and monodisperse nanotubes, which is certainly an oversimplification of the real CNTs-SU8 system.
The effect discussed here is however rather general since it relies only on the enhanced
probability of having nanotubes with short separations. Furthermore, effects from waviness and size polydispersity of nanotubes 
are expected to partially compensate each other, since they affect conductivity in the opposite 
way.\cite{Otten2009,Chatterjee2010,Berhan2007} The effect of correlations (which can be linked to the contact value
of the radial distribution function) on the percolation properties of polydisperse rods has been recently studied in 
Ref.~\onlinecite{Chatterjee2012}.

In conclusion, the practical message of this study is that one can make a well-defined, high quality composite of CNTs and SU8, 
where the conductivity can be fine tuned by adjusting the volume fraction of CNTs or the processing conditions of the composite. 
On the fundamental research side the N-P composite set is especially interesting. It corresponds to the conduction in a liquid-like 
matrix which is well described by tunneling processes between homogeneously dispersed CNTs. On the contrary, the sets of polymerized
composites (P and P-B) follow a percolation-like dependence of the conductivity, which we interpret as due to a reduction of 
the mean nanotube separation induced by polymerization of the SU8 matrix. 
It is worth to point out that our interpretation predicts that the percolation-like behavior of the 
polymerized composites is expected to evolve into a tunneling-like one for $\phi$ lower than $1/\chi$, which is however outside of our concentration range for the P and P-B samples. 
Finally, our results suggest that the polymerized SU8 matrix has an active role in the conduction
process of the composites, and that enhanced conductivity is attained at low CNT concentrations as a result
of an effective CNT ``stickiness'' driven by the polymerization process.

%It is interesting to notice that the criterion for percolation threshold for the N-P composite is observed at $\phi = 0.002$ volume 
%fraction, where $\sigma$ switches over from the tunneling to the percolating regime. One would expect that the same threshold would 
%be observed in the polymerized samples (P and P-B) just normalized by a volume shrinking by $30$ and $45\%$, respectively. 
%Surprisingly, this scaling does not hold. The fact that this threshold is out of our concentration scale for the P and P-B samples, 
%suggests that the polymerized matrix has also an active role in the conduction process of the composites.

Useful discussions with J. Jacimovic, T. Stora and P. van der Schoot are gratefully acknowledged. 
This work is supported by the Swiss National Science Foundation (NSF 113723) and Gersteltec Sarl.


\begin{thebibliography}{99}

\bibitem{Byrne2010} M. T.  Byrne and Y. K. Gun\'ko, Adv. Mater. \textbf{22} , 1672 (2010). 

\bibitem{Brennan2011} L. J. Brennan, M. T. Byrne, M. Bari, and Y. K. Gun\'ko, Adv. Energy Mater.  \textbf{1}, 472 (2011).

\bibitem{Lota2011} G. Lota, K. Fic, and E. Frackowiak, Energ. Environ. Sci. \textbf{4}, 1592 (2011). 

\bibitem{USPatents} US Patent no. 4882245: Photoresist Composition and Printed Circuit Boards and Packages Made Therewith, 
J. D. Gelorme, R. J. Cox, and S. A. R. Gutierrez (1989); US Patent no. 4940651: Method for Patterning Cationic Curable Photoresist, 
L. M. Brown, J. D. Gelorme, J. P. Kuczynski, and W. H. Lawrence (1990).

\bibitem{Teh2005} W. H. Teh, U. D\"urig, U. Drechsler, C. G. Smith, and H.-J. G\"untherodt, J. Appl. Phys. \textbf{97}, 054907 (2005).

\bibitem{Jiquet} 
S. Jiguet, M. Judelewicz, S. Mischler, H. Hofmann, A. Bertsch, and P. Renaud, Surf. Coat. Technol. \textbf{201}, 2289 (2006); 
S. Jiguet, A. Bertsch, M. Judelewicz, H. Hofmann, and P. Renaud, Microelectron. Eng. \textbf{83}, 1966 (2006); 
S. Jiguet, M. Judelewicz, S. Mischler, A. Bertsch, and P. Renaud, Microelectron. Eng. \textbf{83}, 1273 (2006); 
S. Jiguet, A. Bertsch, H. Hofmann, and P. Renaud, Adv. Funct. Mater. \textbf{15}, 1511 (2005); 
S. Jiguet, A. Bertsch, H. Hofmann, and P. Renaud, Adv. Eng. Mater. \textbf{6}, 719 (2004). 

\bibitem{Xu2002} X. Xu, M. M. Thwe, C. Shearwood, and K. Liao, Appl. Phys. Lett. \textbf{81}, 2833 (2002). 

\bibitem{Zhang2003} N. Zhang, J. Xie, M. Guers, and V. K. Varadan, Smart Mater. Struct. \textbf{12}, 260 (2003).

\bibitem{Mionicthesis} M. Mioni\'c, Ph. D. thesis N. 5248, Ecole Polytechnique F\'ed\'erale de Lausanne (EPFL), 2011.

\bibitem{Mionic} M. Mioni\'c, K. Pataky, R. Gaal, A. Magrez, J. Brugger, and L. Forr\'o, 
J. Mater. Chem.  \textbf{22}, 14030 (2012).

\bibitem{PatentLPCM} M. Mioni\'c, A. Magrez, L. Forr\'o, S. M. Jiguet, M. P. Judelewicz and T. Stora, Carbon nanotubes/SU-8 nanocomposites for microfabrication applications, Patent No., WO/2011/061708, 2011. 

\bibitem{Ong2006}
B. H. Ong, X. Yuan, S. Tao and S. C. Tjin, Opt. Lett. \textbf{31}, 1367 (2006).

\bibitem{Feng2002}
R. Feng and R. J. Farris, J. Mater. Sci. \textbf{37}, 4793 (2002).

\bibitem{Stauffer1992} D. Stauffer and A. Aharony, \textit{Introduction to Percolation Theory} (Taylor and Francis, London, 1992).

\bibitem{Ambrosetti2010} G. Ambrosetti, C. Grimaldi, I. Balberg, T. Maeder, A. Danani, and P. Ryser, 
Phys. Rev. B \textbf{81}, 155434 (2010).

\bibitem{Grimaldi2011} G. Ambrosetti, I. Balberg, and C. Grimaldi, Phys. Rev. B \textbf{82}, 134201 (2010); 
C. Grimaldi, Europhys. Lett. \textbf{96}, 36004 (2011).

\bibitem{note1} The factor $2$ appearing in front of Eq.~\eqref{EMA1} has been introduced in order to reproduce
the correct tunneling limit of Eq.~\eqref{EMA3}.

\bibitem{Nigro2012}
B. Nigro, C. Grimaldi, M. A. Miller, P. Ryser, and T. Schilling,
J. Chem. Phys. \textbf{136}, 164903 (2012).

\bibitem{Ambegaokar1971} V. Ambegaokar, B. I. Halperin, and J. S. Langer, Phys. Rev. B \textbf{4}, 2612 (1971).

\bibitem{Otten2009} R. H. J. Otten and P. van der Schoot, Phys. Rev. Lett. \textbf{103}, 225704 (2009); 
J. Chem. Phys. \textbf{134}, 094902 (2011).

\bibitem{Chatterjee2010} A. P. Chatterjee, J. Chem. Phys. \textbf{132}, 224905 (2010); J. Stat. Phys. \textbf{146}, 244 (2012).

\bibitem{Straley1977} J. P. Straley, J. Phys. C \textbf{10}, 3009 (1977).

\bibitem{Berhan2007} L. Berhan and A. M. Sastry, Phys. Rev. E \textbf{75}, 041121 (2007).

\bibitem{Chatterjee2012} A. P. Chatterjee, J. Chem. Phys. \textbf{137}, 134903 (2012).




\end{thebibliography}
\end{document}